\documentclass[aps,twocolumn,epsfig,graphics,showpacs,floatfix]{revtex4}
\usepackage{amsmath,amsfonts,amssymb,graphics,graphicx,epsfig,color,times}
\begin{document}
\title{Mixtures of Bosonic and Fermionic Atoms in Optical Lattices}
\author{Alexander Albus$^{1,2}$, Fabrizio Illuminati$^{2}$, and 
Jens Eisert$^{1,3}$}
\affiliation{\mbox{}$^{1}$Institut f\"{u}r Physik, 
Universit\"{a}t Potsdam, Am Neuen Palais 10,
D--14469 Potsdam, Germany\\
\mbox{}$^{2}$Dipartimento di Fisica, Universit\`{a} di Salerno, 
and Istituto Nazionale per la Fisica della Materia, I--84081 Baronissi
(SA), Italy\\
\mbox{}$^{3}$ Blackett Laboratory, Imperial College London,
Prince Consort Road,
SW7 2BW London, UK}

\date{May 14, 2003}

\begin{abstract}
We discuss the theory of mixtures of Bosonic and Fermionic
atoms in periodic potentials at zero temperature. We derive a
general Bose--Fermi Hubbard Hamiltonian in a one--dimensional
optical lattice with a superimposed harmonic trapping potential.
We study the conditions for linear stability of the mixture 
and derive a mean field criterion for the
onset of a Bosonic superfluid transition.
We investigate the ground state properties
of the mixture in the Gutzwiller formulation of
mean field theory, and present numerical studies of finite systems. 
The Bosonic and Fermionic
density distributions and the onset of quantum phase 
transitions to demixing and to a Bosonic Mott--insulator are
studied as a function of the lattice potential strength.
The existence is predicted of a disordered phase for 
mixtures loaded in very deep lattices. 
Such a disordered phase possessing many degenerate or
quasi--degenerate ground states is related to a 
breaking of the mirror symmetry in the lattice.
\end{abstract}


\maketitle

\section{Introduction}
Recent spectacular progress in the manipulation of
neutral atoms in optical lattices \cite{Greiner1,Greiner2,OrzTuc01}
has opened the way to the simulation of complex
quantum systems of condensed matter physics, such as
high--$T_{c}$ superconductors, Hall systems,
and superfluid ${}^{4}$He, by 
means of atomic systems with perfectly controllable physical
parameters \cite{Anglin}.
Optical lattices are stable periodic arrays of microscopic potentials 
created by the interference patterns of intersecting laser
beams \cite{Jessen1}.
Atoms can be confined to different
lattice sites, and by varying the strength of the periodic potential
it is possible to tune the interatomic interactions 
with great precision. They can be enhanced well into
regimes of strong correlation, even in the dilute limit.
The transition to a strong coupling regime can be realized by increasing
the depth of the lattice potential wells, a quantity that is
directly proportional to the intensity of the laser light. This is
an experimental parameter that can be controlled with
great accuracy.
For this reason, besides the fundamental interest for 
the investigation of quantum phase transitions 
\cite{Sachdev} and other 
basic quantum phenomena 
\cite{JacBru98,OosStr02,RuoDun02,Hofstetter02,Pared02,Recati03,Zwerger03},
optical lattices have become an important practical tool for
applications, ranging from laser cooling \cite{Kerman} to
quantum control and information processing \cite{Jessen2,DuaDemLuk03},
and quantum computation 
\cite{Deutsch,JakBriCirGarZol98,GarCir03,DorFed02,PacKni03}.

The theory of neutral Bosonic atoms in optical lattices has
been developed \cite{JacBru98} by assuming that the atoms are 
confined to the lowest Bloch band of the periodic potential. 
It can then be shown that the system is effectively described by a
single--band Bose--Hubbard model Hamiltonian \cite{Fisher}. 
In such a model the superfluid--insulator transition is predicted
to occur when the on site Boson--Boson interaction energy becomes 
comparable to the hopping energy between adjacent lattice sites. 
This situation can be experimentally achieved by increasing the strength 
of the lattice potential, which results in a strong suppression of the
kinetic (hopping) energy term.
In this way, the superfluid--Mott-insulator quantum phase
transition has been realized by loading 
an ultracold atomic Bose--Einstein condensate in an optical 
lattice \cite{Greiner1}.

The present paper is concerned with the study of
dilute mixtures of interacting Bosonic and Fermionic neutral
atoms subject to an optical lattice and a superimposed
trapping harmonic potential at zero temperature.
We assume the fermions to be identical (for instance 
spin--polarized in a magnetic trap), so that there are
only $s$--wave Boson--Boson and Fermion--Boson contact
interactions present.
We construct an effective single--band Bose--Fermi Hubbard
Hamiltonian, and we determine the ground state energy and
on site density distributions in different mean field 
approximations of increasing complexity.
Our main aim at this level of description is to
determine the basic ground state properties of the
mixture and to study how the 
Bosonic superfluid--insulator transition is influenced by
the presence of the Fermions. 
Besides the study of the latter issue, and the assessment
of the properties of linear stability of the system
against Boson--Fermion demixing, a remarkable finding
of our analysis is that a quantum binary mixture loaded
in a very deep optical lattice allows for a disordered
phase of very many degenerate or quasi--degenerate
ground states separated by very high potential energy
barriers. In the limit of very large lattice potential strengths
the basic mirror symmetry of the optical lattice is broken.

The plan of the paper is as follows: In Section \ref{Model}
we set the notation and derive the Bose--Fermi Hubbard
model Hamiltonian. We then discuss the range of validity 
of the approximations and the assumptions used in the derivation.
In Section \ref{Stability} we introduce some basic mean field
descriptions to study the properties of stability of the mixture 
against phase separation, and we provide a simple analytical 
criterion for the onset of a superfluid phase for the Bosons 
starting from a Mott--insulating ground state.

In Section \ref{Gutzwiller} we present numerical simulations 
for a small number of particles in the framework of the Gutzwiller 
variational ansatz. Two important cases have to be distinguished:
Boson--Fermion repulsive or attractive interaction 
(the Boson--Boson interaction
is taken to be always repulsive). In the first instance, one
can observe a continuous transition to a complete demixing of 
the Fermions from the Bosons, and to a Mott--insulating phase of 
the Bosons, as the strength of the lattice potential is increased.
In the case of attractive Boson--Fermion interactions there
is no transition to demixing or to collapse of the mixture,
while one still observes a Mott--insulating transition for the
Bosons. By studying the behavior of the superfluid order
parameter we show that in both cases the transition takes place
at the same critical value of the lattice potential strength. 
Moreover, due to the strong attraction, the Fermions
and the Bosons tend to form together ordered block--crystalline 
structures at the center of the trap. 

In Section \ref{Degeneracy} we present a numerical analysis that,
although constrained to a small
number of particles (five bosons and five fermions), seems to 
indicate the existence of a rich structure of 
degenerate energy minima for large enough
values of the lattice potential strength. 
Such an energy landscape suggests the 
possible existence of disordered phases of the mixture due to the
delicate interplay between the different physical parameters 
(Boson hopping, Fermion hopping, Boson on site energy, 
Boson--Fermion on site energy), the lattice depth, 
and the symmetries of the problem.
Finally, a summary and an outlook to future research are
shortly discussed in Section \ref{Outlook}.

\section{Model Hamiltonian}
\label{Model}
We start by introducing the 
Hamiltonian for a Bose--Fermi mixture
loaded into optical lattice potentials
and confined by additional, slowly varying,
external (harmonic) trapping potentials. It is given by
\begin{equation}\label{Fully}
\hat{H}=\hat{T}_B+\hat{T}_F+\hat{V}_B
+\hat{V}_F+\hat{W}_{BB}+\hat{W}_{BF},
\end{equation}
\noindent where
\begin{eqnarray}
\hat{T}_{B}&=&
-{\displaystyle \int } d^3{\bf r}\hat{\Phi}^\dagger({\bf r})
\frac{\hbar^2{\bf \nabla}^2}{2m_B}\hat{\Phi}({\bf r}) \; , \\
\hat{T}_{F}&=&
-{\displaystyle \int } d^3{\bf r}\hat{\Psi}^\dagger({\bf r})
\frac{\hbar^2{\bf \nabla}^2}{2m_F}\hat{\Psi}({\bf r}) \; ,
\end{eqnarray}
represent the Boson
and Fermion kinetic energies, respectively, while
\begin{eqnarray}
\hat{W}_{BB}&=&\frac{1}{2}\frac{4\pi^2\hbar^2a_{BB}}{m_B}
\int d^3{\bf r}\hat{\Phi}^\dagger({\bf r}) 
\hat{\Phi}^\dagger({\bf r})
\hat{\Phi}({\bf r})\hat{\Phi}({\bf r}),\\
\hat{W}_{BF}&=&\frac{2\pi^2\hbar^2a_{BF}}{m_R}
\int d^3{\bf r} 
\hat{\Phi}^\dagger({\bf r}) \hat{\Psi}^\dagger({\bf r})
\hat{\Psi}({\bf r})\hat{\Phi}({\bf r}) \, ,
\end{eqnarray}
denote the Boson--Boson and the
Fermion--Boson contact interaction energies. They are
parametrized 
by the Boson--Boson and the Fermion--Boson 
$s$--wave scattering lengths $a_{BB}$ and $a_{BF}$, respectively,
and by the Boson mass $m_{B}$ and the reduced mass
$m_{R} = m_{B}m_{F}/(m_{B} +m_{F})$, where $m_{F}$
denotes the Fermion mass.
The potential energies
\begin{eqnarray}
\hat{V}_{B}&=&\int d^3{\bf r} 
\hat{\Phi}^\dagger({\bf r})
\left(V_B({\bf r})+P_B({\bf r})\right)\hat{\Phi}({\bf r}) \, , \\
\hat{V}_{F}&=&\int d^3{\bf r}
\hat{\Psi}^\dagger({\bf r})
\left(V_F({\bf r})+P_F({\bf r})\right)\hat{\Psi}({\bf r}) \, ,
\end{eqnarray}
are due to the trapping and lattice potentials.
We consider pure magnetic trapping
for Bosons and Fermions, so that Fermions are spin--polarized
and their $s$--wave interaction energy $\hat{W}_{FF}$ 
can be neglected: $\hat{W}_{FF} = 0$.
In the subsequent analysis we will consider the harmonic
approximation of a typical quadrupolar magnetic
field with strong anisotropy
in the transverse directions $y$ and $z$, i.e.,
\begin{eqnarray}
V_{B}({\bf r}) \simeq m_{B}
\omega_{B}^{2}(x^{2} + \lambda^{2}y^{2} +
\lambda^{2}z^{2})/2 \; ,
\end{eqnarray} 
and 
\begin{equation}
V_{F}({\bf r}) \simeq m_{F}
\omega_{F}^{2}(x^{2} + \lambda^{2}y^{2} +
\lambda^{2}z^{2})/2 \; ,
\end{equation}
where $\lambda \gg 1$ is the anisotropy
parameter.
Moreover, if we assume trapping in the same magnetic state
for the Bosons and the Fermions, then
the trapping frequencies are related according to
$\omega_{F}/\omega_{B}=(m_B/m_F)^{1/2}$, so that the two 
potentials coincide: $V_{B}({\bf r}) = V_{F}({\bf r})$.
The ground--state harmonic oscillator lengths, however, 
are different due to the different masses, and also differ 
for the $x$-direction on the one hand and the $y$ and $z$-directions
on the other hand:
\begin{equation}
\ell_{B/F}^\parallel = \sqrt{\hbar/(m_{B/F}\omega_{B/F})}
\end{equation}
in the $x$ direction, and $\ell_{B/F}^\perp = 
\ell_{B/F}^\parallel/\sqrt{\lambda}$ in the $y$ and
$z$ directions.
We next consider a lattice structure for the Bosons and the
Fermions in the $x$--direction,
associated to  the corresponding Bosonic and 
Fermionic one--dimensional
optical lattice potentials $P_{B}(x)$ and $P_{F}(x)$ 
are
\begin{eqnarray}
P_{B}(x) & = & V^{0}_{B}\sin^{2}(\pi x/a) \; , \nonumber \\
P_{F}(x) & = & V^{0}_{F}\sin^{2}(\pi x/a) \; ,
\end{eqnarray}
where $a$ is the lattice spacing associated to the wave
vector $k = \pi/a$ of the standing laser light.
If the lattice potentials are produced by a far off--resonant laser
for both species, the lattice potential strengths 
are equal for both Fermions
and Bosons: $V^{0}_{F} = V^{0}_{B} = V_{0}$, 
and the two optical lattices coincide exactly. 
This is the situation we will always consider
in the following.

In the presence of a strong optical lattice and a sufficiently
shallow external confinement in the $x$ direction, the field operators can
be expanded in terms of the single--particle Wannier functions 
localized at each lattice site $x_i$. Further, the typical
interaction energies involved are normally not strong enough in order
to excite higher vibrational states, and we can retain only the lowest
vibrational state in each lattice potential well both for Bosons 
and Fermions (single--band approximation). 
In case of stronger external confinements, or 
interactions, one should include higher Bloch bands as well
in the expansion of the field operators, 
a case we do not consider in the present context. In the harmonic
approximation, the Wannier functions $w({\bf r})$ factorize in the 
product of harmonic oscillator states in each direction, with the
trapping potential almost constant between adjacent lattice sites.
We then have
\begin{eqnarray}
\hat{\Phi}({\bf r}) & = &
\sum_i\hat{a}_iw^B_x(x-x_{i})w^B_y(y)w^B_z(z) \, , 
\label{expansion1}
\end{eqnarray}
\begin{eqnarray}
\hat{\Psi}({\bf r}) & = &
\sum_i\hat{b}_iw^F_x(x-x_{i})w^F_y(y)w^F_z(z) \, ,
\label{expansion2}
\end{eqnarray}
where $\hat{a}_i$ and $\hat{b}_i$ are respectively 
the Bosonic and Fermionic annihilation operators at the
$i$--th lattice site, $x_{i} = ia$, and the index $i$
runs on positive and negative integers, the origin of the 
lattice being fixed at $i=0$ so that it coincides with the
center (the minimum) of the external trapping potential. 
In each lattice potential well
the Wannier local ground states for Bosons 
and Fermions are Gaussians in the harmonic approximation:
\begin{eqnarray}
w^{B/F}_y(y) & = & \frac{\exp{\left[ 
-y^{2}/2(\ell^\perp_{B/F})^{2} 
\right]}}{\pi^{1/4}(\ell^\perp_{B/F})^{1/2}}
\, , 
\label{Wanniery}
\end{eqnarray}
\begin{eqnarray}
w^{B/F}_z(z) & = & \frac{\exp{\left[ 
-z^{2}/2(\ell^\perp_{B/F})^{2} 
\right]}}{\pi^{1/4}(\ell^\perp_{B/F})^{1/2}}
\, , 
\label{Wannierz}
\end{eqnarray}
and
\begin{equation}
w^{B/F}_x(x-x_{i}) = \frac{\exp{\left[ 
-(x-x_{i})^{2}/2(\ell^{0}_{B/F})^{2} 
\right]}}{\pi^{1/4}(\ell^{0}_{B/F})^{1/2}}
\; , 
\label{Wannierx}
\end{equation}
where
\begin{equation}
\ell_{B/F}^{0} = a/[\pi(V_{0}/E^R_{B/F})^{1/4}] \; ,
\end{equation}
is the width of the harmonic oscillator potential wells
at each lattice site, with
$E^R_{B} = (\pi\hbar)^{2}/2a^{2}m_{B}$ and 
$E^R_{F} = (\pi\hbar)^{2}/2a^{2}m_{F}$ being the Boson 
and Fermion recoil energies, respectively.

\begin{figure}[h]
\begin{center}
\epsfysize=8.5cm
\rotatebox{-90}{\epsfbox{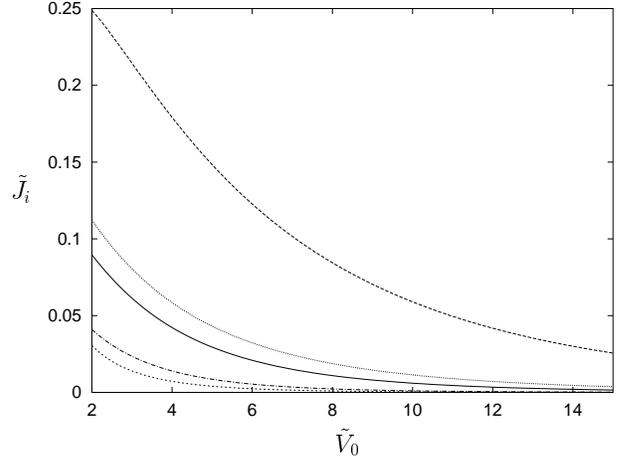}}
\end{center}
\caption{Top to bottom: the Fermion hopping amplitude 
${\tilde{J}}_{F}$ for $m_F/m_B = 0.5$ in the
Gaussian approximation;
the Boson hopping amplitude ${\tilde{J}}_{B}$ from
the exact 1--$D$ Mathieu equation;
the Boson hopping amplitude ${\tilde{J}}_{B}$
in the Gaussian approximation;
the Fermion hopping amplitude for $m_F/m_B=1.5$
in the Gaussian approximation; 
and, for comparison, the overlap integral 
$\langle w(x - x_{i}) | w(x - x_{i + 1}) \rangle $
of adjacent Wannier functions in the Gaussian
approximation. All quantities are dimensionless.}
\label{JBFfig}
\end{figure}

In this paper we will consider the physical situation of
very shallow trapping potentials, such that 
$\ell_{B/F}^{\parallel} \gg aN_{B/F}$ and consequently
local density approximation (LDA) can be applied in the 
study of the ground--state properties of the system.
Therefore, when exploiting the Wannier function expansions 
(\ref{expansion1}) and (\ref{expansion2}) to map the full Hamiltonian 
(\ref{Fully}) into its lattice version, we 
discard all terms that are of order 
$(aN_{B/F}/\ell_{B/F}^{\parallel})^{2}$ or of higher powers
of it. Otherwise, nonlocal effects caused by the trapping
potential, like site--dependent hopping terms, have
to be considered \cite{Inpreparation}. Finally,
this approximation scheme leads to the following
Hubbard--type Hamiltonian:
\begin{eqnarray}\label{Hubbard}
\hat{H} & = & -\frac{1}{2}\sum_{i}
\left(J_B\hat{a}_{i+1}^\dagger\hat{a}_i + 
J_F\hat{b}_{i+1}^\dagger\hat{b}_i+\mbox{H. c.}\right) 
\nonumber \\
& + & \frac{U_{BB}}{2}
\sum_i\hat{n}_{B}^{(i)}(\hat{n}_{B}^{(i)} - 1)
+ U_{BF}\sum_i\hat{n}_{B}^{(i)}\hat{n}_{F}^{(i)}
\nonumber \\
& + & \sum_i V_{B}^{(i)}\hat{n}_{B}^{(i)}
+ \sum_i V_{F}^{(i)}\hat{n}_{F}^{(i)} \\
& + & \hbar\left(\lambda \omega_B + \omega_{B}^{0}/2 \right)
\hat{N}_{B} +
\hbar \left(\lambda \omega_F + \omega_{F}^{0}/2 \right)
\hat{N}_{F} \; .
\nonumber
\end{eqnarray}
The first line in the above Bose--Fermi Hubbard Hamiltonian
describes independent nearest--neighbor hopping of Bosons
and Fermions, with amplitudes $J_{B}$ and $J_{F}$, respectively.
The terms in the second line describe Boson--Boson on
site repulsion (with $U_{BB} > 0$) and Boson--Fermion on
site interaction. This interaction can be repulsive or attractive, 
depending on
the sign of $U_{BF}$. 
The third line describes the 
energy offset at each lattice site due to the 
$x$ component of the external trapping potentials 
$V_{B/F}({\bf r})$, 
and the last line contains the overall constant zero--point 
energy terms due to the $y$ and $z$ components of 
$V_{B/F}({\bf r})$
and to the lattice potential $P(x)$. The on site interaction and offset
energy terms are simple functions of the on site Boson and 
Fermion occupation number operators $\hat{n}_{B}^{(i)} = 
\hat{a}_{i}^\dagger\hat{a}_{i}$ and $\hat{n}_{F}^{(i)} =
\hat{b}_{i}^\dagger\hat{b}_{i}$, while the zero--point
energy terms are proportional to the 
total particle number operators $\hat{N}_{B} =
\sum_{i}\hat{a}_{i}^\dagger\hat{a}_{i}$ and
$\hat{N}_{F} =
\sum_{i}\hat{b}_{i}^\dagger\hat{b}_{i}$.
The frequency 
\begin{equation}
\omega_{B/F}^{0} = 
\hbar/[(\ell_{B/F}^{0})^{2}m_{B/F}]
\end{equation}
fixes the
Bosonic and Fermionic harmonic oscillations in each
lattice well.
The relevant parameters entering in the Hamiltonian are 
the on site values of the trapping harmonic potential
\begin{eqnarray}
V_{B/F}^{(i)} & = & \frac{m_{B/F}}{2}\omega_{B/F}^{2}x_{i}^{2} \; ,
\end{eqnarray}
the nearest--neighbor hopping amplitudes between adjacent
sites $x_{i}$ and $x_{i+1}$ for Bosons and Fermions
\begin{eqnarray}
J_{B/F}&=&\int dx\;
w^{B/F}_x(x-x_{i})\left[ -\frac{\hbar^2}{2m_{B/F}}\frac{d^2}{dx^2}
\right. \nonumber \\
& + & 
V_{0}\sin^{2}\left( \pi \frac{x}{a} \right) \bigg] 
w^{B/F}_x(x-x_{i+1}) \; ,
\end{eqnarray}
the strength of the on site repulsion energy between two 
Bosonic atoms at the same lattice site
\begin{eqnarray}
U_{BB}&=&\frac{4\pi\hbar^2a_{BB}}{m_B}\int dx\;(w_x^B(x-x_{i}))^{4}
\nonumber\\& \times &
\int dy\;(w_y^B(y))^{4}\int dz\;(w_z^B(z))^{4} \; ,
\end{eqnarray}
and the strength of the on site interaction energy (either 
repulsive or attractive) between a Bosonic and a Fermionic
atom at the same lattice site 
\begin{eqnarray}
U_{BF} & = & \frac{2\pi\hbar^2a_{BF}}{m_R}
\int dx \left[ w_x^B(x-x_{i}) w_x^F(x-x_{i})\right]^{2}\\
&\times&\int dy \left[ w_y^B(y) w_y^F(y) \right]^{2} \int 
dz \left[ w_z^B(z) w_z^F(z) \right]^{2} \; .\nonumber
\end{eqnarray}
In typical situations we may neglect next--to--nearest 
neighbor hopping amplitudes and nearest--neighbor interaction
couplings
that are usually some orders of magnitude smaller, so that
the Hamiltonian (\ref{Hubbard}) provides a rather accurate
model for the dynamics of a Bose--Fermi mixture with 
three--dimensional scattering
in a one--dimensional periodic potential.
Terms involving nearest--neighbor interaction strengths
and/or next--to--nearest neighbor hopping amplitudes
can become relevant and need to be included, e.g., when 
considering phonon exchange between Fermions, and this
would lead to a Bose--Fermi analog of the so--called
extended Hubbard models.
To evaluate estimates for the parameters entering the Bose--Fermi
Hubbard Hamiltonian (\ref{Hubbard}) 
using Eqns.\ (\ref{Wanniery}),(\ref{Wannierz}) and (\ref{Wannierx}),
we will set the Boson recoil energy
$E_{B}^{R} = \hbar^2\pi^2/(2m_{B}a^{2})$
as the unit of energy. We then introduce the dimensionless
quantity ${\tilde{V}}_{0} = V_{0}/E_{B}^{R}$, and,
analogously, the dimensionless quantities
${\tilde{U}}_{BB}$, ${\tilde{U}}_{BF}$,
${\tilde{V}}_{B}^{(i)}$, ${\tilde{V}}_{F}^{(i)}$, 
${\tilde{J}}_{B}$, and ${\tilde{J}}_{F}$. 
We then have
\begin{eqnarray}
{\tilde{U}}_{BB} & = & \sqrt{\frac{8}{\pi^{3}}}
\, \frac{a_{BB}\, a}{(\ell_B^\perp)^2} \, 
{\tilde{V}}_{0}^{1/4} \; , \label{UBB} \\
&& \nonumber \\
{\tilde{U}}_{BF} & = & 
\sqrt{\frac{8}{\pi^{3}}}\left( 1 + \frac{m_{B}}{m_{F}}\right)
\frac{a_{BF} \, a}{(\ell_B^\perp)^2+(\ell_F^\perp)^2}
{\tilde{V}}_{0}^{1/4} \, , \label{UBF} 
\end{eqnarray}

\begin{eqnarray}
{\tilde{V}}_{B}^{(i)} = 
\frac{i^{2}}{\pi^2(\ell_{B}^{\parallel}/a)^4},\quad
{\tilde{V}}_{F}^{(i)} = \frac{m_{B}}{m_{F}} \,  
\frac{i^{2}}{\pi^2(\ell_{F}^{\parallel}/a)^4} \; ,
\end{eqnarray}

\begin{eqnarray}
{\tilde{J}}_{B} & = &
\left( \frac{\pi^{2}}{4} - 1 \right) 
{\tilde{V}}_{0} 
\exp{ \left[ -\frac{\pi^{2}}{4}\sqrt{{\tilde{V}}_{0}} \, \right] }
\; 
,\label{JB}\\
&& \nonumber \\
{\tilde{J}}_{F} & = & 
\left( \frac{\pi^{2}}{4} - 1 \right) 
{\tilde{V}}_{0}
\exp{ \left[ -\frac{\pi^{2}}{4}\sqrt{\frac{m_{F}}{m_{B}}{\tilde{V}}_{0}} 
\, \right] } \; .
\label{JF}
\end{eqnarray}
In FIG. \ref{JBFfig} we show the dependencies of the 
these parameters on the potential strength ${\tilde{V}}_{0}$
(compare also Ref. \cite{DuaDemLuk03}).
For reference we have included as well the overlap integral
\\
\noindent
$\langle w(x - x_{i}) | w(x - x_{i + 1}) \rangle $
of adjacent Wannier functions. The overlap is negligible 
but for very small values of the potential strength, confirming
that terms of the order of the overlap integral can be neglected
in the Hamiltonian.
The Gaussian approximation holds rather well as can be seen
by comparing the associated Bosonic hopping amplitude $J_{B}$ 
with the one obtained by using the exact 
1--$D$ Mathieu equation \cite{Zwerger03}.

Besides the conditions mentioned earlier, all the expressions derived 
in the present section are justified under the following circumstances.
First of all, we must require that the two--body scattering processes
are not influenced by the confinements, a condition that is guaranteed
if the lengths of the confining and lattice potentials in all directions 
are much larger than the Boson--Boson and Fermion--Boson scattering
lengths. Next, the single--band structure of the lattice Hamiltonian
is assured if the lattice spacing $a$ is much greater than the 
harmonic confinements in each direction at all lattice sites.
On the other hand, in this limit the harmonic approximation 
for the Wannier functions at each lattice well 
is automatically satisfied.
Finally, as mentioned earlier, the assumption of a slowly varying confining 
potential such that LDA is applicable leads to the condition
$\ell_{B/F}^{\parallel} \gg aN_{B/F}$.
We can summarize all the above conditions with the following chain 
of inequalities:
\begin{equation}
\{ |a_{BF}|,a_{BB} \} \ll 
\{ \ell_{B/F}^{0}, \ell^\perp_{B/F}\} \ll 
a \ll \ell^{\parallel}_{B/F}/N_{B/F} \, .
\label{chain}
\end{equation}

Our model is for some aspects unrealistic, since
in present experimental situations the transverse
confinements cannot be made very strong. Therefore
a multi--band structure can appear with several
radial states being occupied, as reported in a
recent experiment by the Florence group on Bose--Fermi 
mixtures in a  $1$--$D$ optical lattice
\cite{GiovanniModugno}.
 
\section{Phase stability and the superfluid transition}
\label{Stability}
In this section we investigate the zero temperature
ground state properties of the system in a mean field approximation.
In the following we will adopt a grand-canonical description
through the Hamiltonian 
\begin{equation}
\hat{K} = \hat{H} - \mu_{B}\hat{N}_{B} - 
\mu_{F}\hat{N}_{F} \; ,
\end{equation}
where $\mu_B$ and $\mu_F$ are the Bosonic and Fermionic
chemical potentials.
According to the Hohenberg--Kohn theorem, the ground
state energy 
\begin{equation}
E = \langle \Psi_0 |\hat{K}| \Psi_0 \rangle
\end{equation}
is a functional of the on site Bosonic and Fermionic densities
$n_B^{(i)} = \langle \hat{a}_i^\dagger\hat{a}_i \rangle$
and $n_F^{(i)} = \langle \hat{b}_i^\dagger\hat{b}_i \rangle$,
where the expectation values are taken with respect to
the ground state with state vector $| \Psi_0 \rangle$.
We decompose the functional $E$ according to
\begin{eqnarray}
E = E_B + E_F + E_{BF} - \mu_B\sum_{i}n_B^{(i)} 
-\mu_F \sum_in_F^{(i)} \, ,
\end{eqnarray}
where $E_B$ is the energy contribution depending only
on the Boson parameters $J_{B}$, $U_{BB}$, $V_{B}^{(i)}$;
$E_F$ is the energy depending only on the Fermion parameters;
and $E_{BF}$ is the term due to Boson--Fermion interactions.
We treat this latter term in mean field approximation: 
neglecting exchange correlation effects:
\begin{equation}
E_{BF} = U_{BF}\sum_i n_B^{(i)}n_F^{(i)} \; .
\end{equation}

Exchange correlation effects have been recently studied for 
the case of homogeneous mixtures in the continuum 
\cite{Uns,Giorgini}. 
For the Fermion energy $E_F$, we take the energy of the
noninteracting homogeneous system and exploit local density
approximation (LDA) on it,
\begin{eqnarray}
E_F = -\frac{2J_F}{\pi}\sum_{i}\sin(\pi n_F^{(i)}) 
+ \sum_i V_F^{(i)}n_F^{(i)} \; .
\end{eqnarray} 
This approximate description of the Fermions is
well justified in the presence of a slowly varying
trapping potential (so that LDA can be applied), 
when there are no direct interactions
among the Fermions (as in our case), and moreover 
when one can neglect induced phonon--mediated 
self--interactions due to the 
presence of the Bosons. Therefore, in this situation,
the nontrivial features of different quantum phases
will regard only the Bosonic sector and not the Fermionic one.
However, the presence of the Fermions will indirectly
contribute to the properties of the different Bosonic phases,
and this is the subject that we will study in the
following. 

In order to find an expression for the Boson energy
$E_{B}$ we will proceed in steps of increasing accuracy.
First we perform a very simple mean field analysis in two
extreme limits: a completely superfluid Boson ground state and a 
totally Mott--insulating Boson ground state. In the latter case
we will provide a simple criterion for stability of the
mixture against demixing.
Next, we will perform a perturbation expansion around the 
Mott--insulating Boson ground state to recover perturbatively
the phase boundary against transition to superfluidity.
Finally, in the next section, we will study the ground
state properties of the mixture using a Gutzwiller
ansatz for the Bosons capable of describing the
intermediate regimes between the insulating and 
superfluid Bosonic phases.

We first consider 
the Bosons to be superfluid.
In this regime the chemical potential and the
number of particles in a homogeneous system are 
related, to lowest 
order in $U_{BB}$, via \cite{OosStr02}:
\begin{eqnarray}
\mu_B=U_{BB}n_0 - 2J_B \; ,
\end{eqnarray}
where $n_0$ is the density of condensed Bosons. 
Additionally, for very weak interaction
$n_0\approx n_B$. Exploiting this result in LDA
and using the mean field expression for the Bose--Fermi
interaction energy we can then write for the inhomogeneous
Bose--Fermi mixture at a given lattice site:
\begin{eqnarray}
U_{BB}n_B^{(i)} = \mu_B+2J_B-V_B^{(i)}-U_{BF}n_F^{(i)} \; .
\label{nbsf}
\end{eqnarray}
Next, we consider the case of a Mott--insulating Bosonic
phase. To lowest order in $J_{B}$ we neglect the 
kinetic term altogether. 
Then it is easily
shown that the relation between the Bosonic chemical potential and
the Bosonic density for a homogeneous system is given by
\begin{eqnarray}
\mu_B=U_{BB}n_B-U_{BB}/2 \; .
\end{eqnarray}
Exploiting LDA as before, we have in the inhomogeneous case
at a given lattice site:
\begin{eqnarray}
U_{BB}n_B^{(i)}=\mu_B+U_{BB}/2-V_B^{(i)}-U_{BF}n_F^{(i)} \; .
\label{nbmi}
\end{eqnarray}
Comparing Eqns. (\ref{nbsf}) and (\ref{nbmi}), we observe the same
behavior of the on site 
density profiles but for a constant correction to the 
Boson chemical potential depending whether
the Bosons are in a 
superfluid or in a Mott--insulating state.
Finally, differentiating the energy functional with respect to the 
on site populations of the Fermions,  
we determine the associated density field and the
set of coupled equations describing the ground state
of the mixture at any lattice site, 
\begin{eqnarray}
U_{BB}n_B^{(i)}&=&\mu'_B-V_B^{(i)}-U_{BF}n_F^{(i)},
\label{BFdensities1}
\end{eqnarray}
\begin{eqnarray}
-2J_F\cos(\pi n_F^{(i)})&=&\mu_F-V_F^{(i)}-U_{BF}n_B^{(i)},
\label{BFdensities2}
\end{eqnarray}
where $\mu'_B$ is the proper expression of the Boson
chemical potential according to whether the Bosons are
in the Mott--insulating  or superfluid regime.
These equations are valid at a given lattice site $i$ if  
$\mu'_B-V_B^{(i)}-U_{BF}n_F^{(i)}>0$, otherwise one must set
$n_B^{(i)}=0$. 
On the other hand, if 
\begin{equation}
(\mu_F-V_F^{(i)}-U_{BF}n_B^{(i)})/2J_F<0
\end{equation} 
we must impose $n_F^{(i)}=0$ at the given lattice site,
while $n_F^{(i)}=1$ must be imposed when
$(\mu_F-V_F^{(i)}-U_{BF}n_B^{(i)})/2J_F>1$.
These expressions are the lattice analogs of the Thomas--Fermi
description of Boson--Fermion mixtures in the continuum.
We remark that in the Mott--insulating regime the Boson
on site populations $n_{B}^{(i)}$ must be rounded off
to the integer closest to the solutions of 
Eqns.~(\ref{BFdensities1})--(\ref{BFdensities2}).

In the Mott--insulating regime we can determine a criterion of 
linear stability against phase separation of
the two species if we expand the energy functional 
$E$ to second order in the small density variations
$\delta n_{B/F}^{(i)}$
around the minimum provided by the solution of Eqns. 
(\ref{BFdensities1})--(\ref{BFdensities2}):
\begin{eqnarray}
\delta^{2} E & = & \frac{1}{2} 
\sum_i\left(\begin{array}{c}\delta n_B^{(i)}\\\delta n_F^{(i)}
\end{array}\right)\cdot \left[
\left(\begin{array}{cc}U_{BB}&U_{BF}\\U_{BF}&2\pi 
J_F\sin(\pi n_{F}^{(i)})\end{array} \right) \right.
\nonumber \\ 
&& \nonumber \\
&& \left. \times \left(\begin{array}{c}\delta n_B^{(i)} \\
\delta n_F^{(i)}\end{array}\right) \right] \; .
\end{eqnarray}
This quadratic form is positive at a given site $i$ if
and only if
\begin{eqnarray}
2\pi J_F\sin(\pi n_{F}^{(i)})U_{BB} > U^2_{BF} 
\label{stability}
\end{eqnarray}
and $2\pi J_F\sin(\pi n_{F}^{(i)}) + U_{BB} \geq 0$.
This last condition is always satisfied 
for $U_{BB} > 0$ and identical Fermions.
If this is not the case for every site $i$, then
the ground state is not stable against demixing.
This result is similar to that recently obtained 
for a mixture of two different Boson species on
a lattice \cite{CheWu03},
which states that the mixture is stable if
$U_{1}U_{2} > U_{12}^{2}$, where $U_{1}$ and $U_{2}$ are
the Boson--Boson interaction strengths of species $1$
and $2$ respectively, and $U_{12}$ is the interspecies
coupling. The form of expression~(\ref{stability})
then suggests that the Pauli on site energy 
$2\pi J_F\sin(\pi n_{F}^{(i)})$
has the meaning of a density--dependent interaction strength.
A similar correspondence was previously pointed out for
homogeneous Bose--Fermi mixtures in the continuum
\cite{Viverit}.

Introducing a perturbation expansion with respect to $J_{B}$
around the Mott--insulating ground state we can recover
the zero--temperature phase transition to the superfluid
phase. The reverse, i.e., to build a perturbative
expansion in powers of $U_{BB}$ around the superfluid ground
state fails to describe the transition to a Mott insulator,
as pointed out in Ref. \cite{OosStr02} for the pure Bose case.
We follow the procedure adopted in Ref. \cite{CheWu03} for the
two--component Boson mixture, with the due modifications for the
present case of a Boson--Fermion mixture, 
by treating the Bosonic kinetic (hopping) 
term as the perturbation with respect to the Bosonic Mott--insulating
ground state. This scheme was first introduced for one--component
Bose systems in Refs. \cite{OosStr02,Freericks,SheKri93}.
We proceed by expanding the ground state energy with respect to the 
(local) Bosonic superfluid parameter $\psi^{(i)} = \langle a_{i} \rangle$. 
At the phase boundary between a Mott insulator (MI) and 
a superfluid (SF) the expansion coefficients
must vanish, yielding the following criterion for the onset of
the transition to the (local) SF state:
\begin{eqnarray}
&&U_{BB}(2n_B^{(i)}-1)-2J_B
\nonumber\\
&&-\left({U^2_{BB}-4U^2_{BB}(2n_B^{(i)}+1)+4J^2_B}\right)^{1/2}
\nonumber\\
&&<\mu_B-V_B^{(i)}-U_{BF}n_F^{(i)}
\nonumber\\
&&<U_{BB}(2n_B^{(i)}-1)-2J_B
\nonumber\\
&&+\left({U^2_{BB}-4U^2_{BB}(2n_B^{(i)}+1)+4J^2_B}\right)^{1/2}.
\end{eqnarray} 
The minimum value of $U_{BB}/J_B$, where a MI phase can exist,
 is given by the condition
\begin{equation}
U_{BB}/J_B=4n_{B}^{(i)}+2+2\sqrt{(2n_{B}^{(i)}+1)^2-1} \; ,
\end{equation}
and it involves the Fermionic sector indirectly through the dependence
of $n_{B}^{(i)}$ on the Fermionic parameters and density distributions
provided by Eqns.~(\ref{BFdensities1})--(\ref{BFdensities2}). 
Apart from this important modification, the phase diagram, 
at this level of approximation, is analogous to that of a 
one--component Bose system.

\section{Number-Conserving Gutzwiller Ansatz and Numerical Analysis}
\label{Gutzwiller}
The simplest ansatz for the Boson ground state being capable of
describing both the SF and the MI phases is the 
Gutzwiller Ansatz, which contains the mean field approximations
previously discussed as special cases.
It consists of factorizing the amplitudes of superpositions of all
possible Fock states consistent with a fixed
number of Bosons $N_B$, in the following way \cite{KrauCaf91}:
\begin{eqnarray}
|\Psi\rangle_B&\longmapsto&
\sum_{\sum_j n_j=N_B} 
\prod_i f_{n_i}^{(i)}\frac{(\hat{a}^\dagger_i)^{n_i}}{\sqrt{n_i!}}
|0\rangle \; .
\end{eqnarray}
In this way correlations between the Bosons under particle
exchange are included in a natural way. 
Using the number--conserving Gutzwiller ansatz in the 
determination of the energy functional, while keeping the same 
approximations previously introduced for the Boson--Fermion
interaction and the Fermion energy, 
the total ground state energy reads
\begin{eqnarray}
E = E_B+E_F+U_{BF}\sum_i n_B^{(i)}n_F^{(i)},
\end{eqnarray}
where the subsidiary conditions ensuring particle number 
conservation are
\begin{eqnarray}
\sum_i n_B^{(i)}&=&\sum_i\langle 
\hat{a}_i^\dagger\hat{a}_i\rangle=N_B,\label{numberB}\\
\quad
\sum_i n_F^{(i)}&=&\sum_i\langle 
\hat{b}_i^\dagger\hat{b}_i\rangle=N_F.\label{numberF}
\end{eqnarray}
The Boson energy contribution is now
\begin{eqnarray}
E_B&=&-\frac{1}{2} J_B \left(
\sum_i {\psi^{(i+1)}}^\ast\psi^{(i)} + \mbox{C. c.} \right)
\nonumber\\
& + & \frac{U_{BB}}{2}(\sigma_B^{(i)}-n_B^{(i)})
+V_B^{(i)}n_B^{(i)} \; ,   
\label{Efunctional}
\end{eqnarray}
and the Bosonic observables are related to the Gutzwiller amplitudes by
\begin{eqnarray}
n_B^{(i)}
&=&\sum_{n=0}^\infty  n (f_n^{(i)})^2,\\
\sigma_B^{(i)}&=&\langle \hat{a}^\dagger_i
\hat{a}_i\hat{a}^\dagger_i\hat{a}_i\rangle
=\sum_{n=0}^\infty n^2 (f_n^{(i)})^2,\\
\psi^{(i)}
&=&\langle \hat{a}_i\rangle = 
\sum_{n=0}^\infty\sqrt{n+1}f_n^{(i)}f_{n+1}^{(i)}\; .
\end{eqnarray}
Moreover, we must impose the natural constraints that
\begin{eqnarray}
	\sum_{n=0}^\infty (f_n^{(i)})^2&=&1,\label{Prob}\\
	0\leq n_F^{(i)}&\leq& 1,\label{Pauli}
\end{eqnarray}
for each lattice site $i$, 
reflecting the fact that the Gutzwiller amplitudes 
form a probability distribution
for each lattice site, and that the on site Fermion 
occupation number cannot exceed one.

To identify the ground state amounts to solving a constrained optimization
problem: one has to minimize the  energy functional 
(\ref{Efunctional}) subject to the
constraints given by Eqns. (\ref{numberB}) and (\ref{numberF}), together
with Eqns. (\ref{Prob}) and (\ref{Pauli}). 
\begin{figure}[h]
\epsfxsize=4.5cm\epsfbox{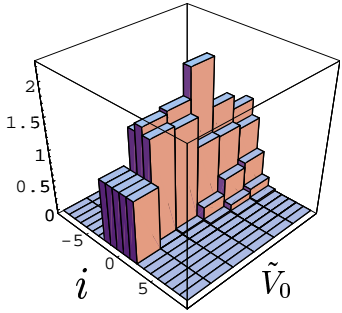}
\caption{On site Bosonic densities for a Bose--Fermi 
repulsion $a_{BF}=0.04$, as a function of the lattice
potential strength. In this figure 
-- as well as in the following figures --
$\tilde{V_0}$ runs from 1 to 8.}
\label{Bdensityrep}
\end{figure}
We have solved the problem numerically for a small system of ten
particles (five Bosons and five Fermions). The first observation
is that the optimization problem is not a convex optimization
problem. Hence, one has to expect several 
local, ``poorer'' extrema in addition to the 
(not necessarily unique) global one.
The numerical solution of this optimization problem 
has been performed first using a simulated annealing method
\cite{KirGelVec83} with an appropriate logarithmic 
annealing schedule. The quadratic constraints 
(\ref{Prob}) and (\ref{Pauli}) have been incorporated
in a dynamical penalty formulation (see, e.g., Ref.\ \cite{WahWan99}).
Finally, for the local refinement the
Nelder--Mead downhill simplex method \cite{NelMea65}
has been applied. 

\begin{figure}[h]
\epsfxsize=4.5cm\epsfbox{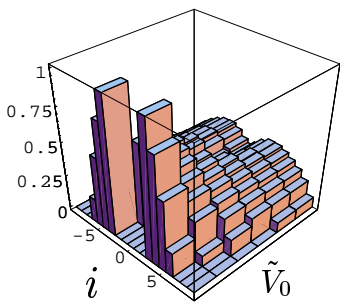}
\caption{On site Fermionic densities for a Bose--Fermi repulsion 
$a_{BF}=0.04$, as a function of the lattice potential strength.}
\label{Fdensityrep}
\end{figure}

In FIG.\ \ref{Bdensityrep} we show the change of the on site Bosonic  
densities with increasing lattice potential strength $\tilde{V}_{0}$
for a system of five Bosons and five Fermions with moderate repulsive
Boson--Fermion interaction.
We note from FIG.\ \ref{Bdensityrep} that as the strength
of the lattice potential increases the Bosons go in a
complete Mott--insulating phase, forming a block crystalline
configuration around the center of the trap (which coincides
with the origin of the optical lattice) with exactly one
Boson per lattice site.
The corresponding on site 
Fermionic densities are plotted in FIG.\ \ref{Fdensityrep}.
From both figures we can see that, 
if $U_{BF} > 0$, by increasing the 
lattice potential strength the system eventually undergoes
simultaneously a Boson MI transition and  
complete phase separation, in accordance with
Eqn. (\ref{stability}) along with Eqns. (\ref{UBB}) -- (\ref{JF}).

\begin{figure}[h]
\epsfxsize=4.5cm\epsfbox{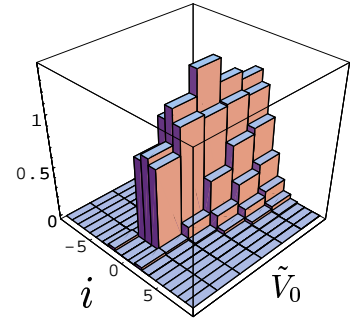}
\caption{The Bosonic superfluid on site order parameter 
for a Bose-Fermi repulsion $a_{BF}=0.04$, as a function
of the lattice potential strength.}
\label{BSFrep}
\end{figure}

The local Bosonic superfluid parameter $\psi^{(i)}$ 
for the same physical situation is shown in 
FIG.\ \ref{BSFrep}. Although we are dealing with
a finite system, we can already see a rather clear
signature of the onset of a phase transition 
to a Mott--insulator regime when the superfluid 
parameter suddenly drops to very low values at an
approximate critical lattice potential 
strength $\tilde{V}_{0}^{c} \simeq 7$.

\begin{figure}[h]
\epsfxsize=4.5cm\epsfbox{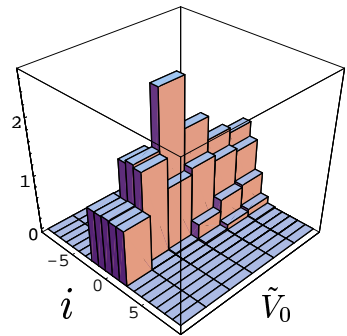}
\caption{On site Bosonic densities for a Bose--Fermi 
attraction $a_{BF}=-0.04$, as a function of the lattice
potential strength.}
\label{BDattract}
\end{figure}

\begin{figure}[h]
\epsfxsize=4.5cm\epsfbox{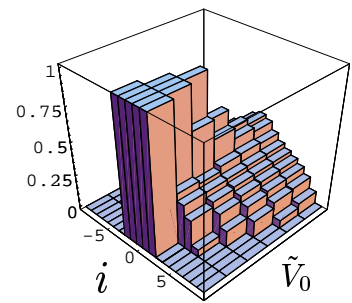}
\caption{On site Fermionic densities for a Bose--Fermi attraction 
$a_{BF}=-0.04$, as a function of the lattice potential strength.}
\label{FDattract}
\end{figure}

\begin{figure}[h]
\epsfxsize=4.5cm\epsfbox{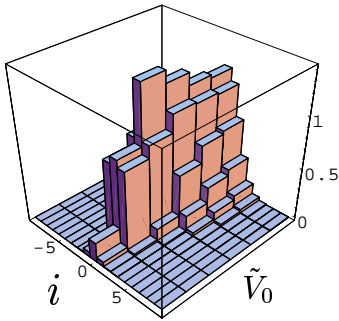}
\caption{The Bosonic superfluid on site order parameter
for a Bose--Fermi attraction $a_{BF}=-0.04$, as a function
of the lattice potential strength.}
\label{BSFattract}
\end{figure}

We next consider the ground--state properties in the case
of an attractive Boson--Fermion interaction. 
Because of the strong attraction with growing lattice depth,
the Fermions follow the Bosons in building a sharp crystalline
block around the center of the trap, as can be seen from
FIGs. \ref{BDattract} and \ref{FDattract}.
We cannot expect in this case 
to observe a simultaneous mean field collapse
like the one predicted for a trapped Bose--Fermi mixture
in the continuum \cite{Roth,Cazalilla} (for the effects beyond mean field
see \cite{Uns2}), as this possibility is forbidden 
in a single--band approximation.
Finally, we consider the behavior of the Bosonic superfluid
on site parameter in the case of a Boson--Fermion attractive
interaction.
Comparing FIG.\ \ref{BSFattract}
with FIG.\ \ref{BSFrep}, we see that the transition to
a Mott insulating phase for the Bosons takes place at the same 
lattice potential strength, irrespectively of the repulsive or
attractive nature of the Boson--Fermion interaction.
This finding confirms the results of the mean field analysis
presented in the previous Section.

\section{Mirror symmetry breaking and transition to degeneracy}
\label{Degeneracy}
The above optimization problem associated with
the constrained minimization of the energy is not convex, 
hence there can be many local minima in addition to
the global one. 
However, even the ground state may be approximately or 
exactly degenerate. In fact,
this is what happens in the case of Boson--Boson
and Boson--Fermion repulsion 
for large values of the lattice potential strength 
$\tilde V_{0}$. As $\tilde V_{0}$ grows, it becomes eventually
energetically more favorable for the bosons to be arranged
in single--particle occupancy of the available sites around
the center of the external trap. The 
Bosonic and Fermionic on--site occupation numbers can
only assume the values $0$ or $1$, and a definite 
Boson--Fermion symmetry is established in the Bose--Fermi
Hubbard Hamiltonian assuming that the on site Fermionic
and Bosonic trapping potentials coincide. 

\begin{figure}[h]
\epsfxsize=4.5cm\epsfbox{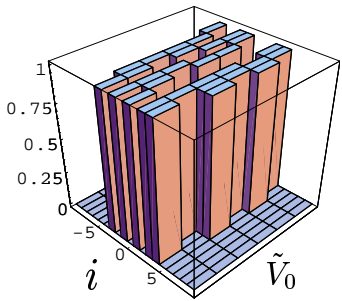}
\caption{The disordered pattern of Bosonic ground--state
distributions for repulsive Boson--Boson and Boson--Fermion
interactions for large values of ${\tilde{V}}_0$ around 
${\tilde{V}}_{0} = 50$.}
\label{Disorder}
\end{figure}

A similar transmutation of Bosons into Fermions in strong
optical lattices has been pointed out by Paredes and Cirac
in a recent paper \cite{Pared02}. They consider a model
of pure Bosons in an optical lattice and show that in the
limit of very strong Boson--Boson on site interaction, the 
Bosonic operators can be mapped into Fermionic operators
by means of the well known Jordan--Wigner transformation.
Let us consider what happens in the case of a Boson--Fermion 
mixture. As the lattice strength grows, 
configurations of lowest energy that are mirror--symmetric 
with respect to the center of the lattice, like e.g. 
those of FIGs. 2 and 3, become approximately 
energetically equivalent
to other symmetric configurations (e.g., a checkerboard of
alternating Bosons and Fermions with one particle per
lattice site), as well as to nonsymmetric configurations 
(like a succession of four Fermions followed by five
Bosons and then a last Fermion, again with one particle
per lattice site), and mirror symmetry breaking takes
place.
We may thus consider
sequences of energy functionals with 
increasing lattice potential strengths $\tilde V_{0}$. For each 
value of $\tilde V_{0}$, one may identify a ground state. 
Then, the difference in energy of this ground state to 
those states that can be 
obtained by interchanging the role of Fermions and Bosons will 
converge to zero as $\tilde V_{0}$ grows. 
The Boson hopping contribution will become negligible,
whereas the behavior of $\tilde U_{BB}$ will
enforce the mean Bosonic on site occupation number 
to be at most one. Hence, for
each lattice site, the constraints on the Boson and Fermion 
occupation numbers become identical (at most one Boson
or one Fermion  per lattice site). Notice that the
suppression of the hopping terms is exponential. Moreover,
since $\tilde V_{B}^{(i)}=\tilde V_{F}^{(i)}$ 
for all lattice sites $i$, the larger the value of
$\tilde V_{0}$, the more symmetric is the role of Bosons and 
Fermions. 
There are then
many ground states that are degenerate in energy with
respect to any permutation of lattice sites -- 
as long as all particles 
are located around the minimum of the 
confining external potential $\tilde V_{B}^{(0)} 
= \tilde V_{F}^{(0)} = 0$. 
These degenerate configurations will be given by all possible  
symmetric and nonsymmetric Fermion and Boson distributions in a
region around the center of the lattice, with every site 
of the region occupied by one and only one particle. 
Such possible configurations are for example checkerboard 
alternating patterns of Bosons and Fermions, or Mott
Bosonic central configurations with Fermionic wings on the sides,
or consecutive block crystalline arrangements of variable
length of Bosons and Fermions. 
In brief, while the Hamiltonian formally retains its mirror
symmetry under reflection of the lattice around its center,
the degenerate ground states need not, and spontaneous
mirror symmetry breaking occurs. At the same time complete
Boson--Fermion exchange symmetry sets on.  
No ground state is {\it a priori} favored compared to 
any other: any random pattern of consecutive Bosons and Fermions 
located around 
the minimum of the external trapping potential is a 
legitimate ground state. 
FIG.\ \ref{Disorder} shows
representative on site Bosonic densities in the regime of large 
values of $\tilde V_{0}$ around ${\tilde{V}}_{0} = 50$
for the case of Boson--Boson and Boson--Fermion repulsion
in a system composed of five Bosons and five Fermions: at each
value of the lattice potential strength, a particular state
is selected from the set of those with same energy.
Each vanishing value of the 
on site Bosonic density means that exactly one 
Fermion has filled that particular lattice site.
The large value chosen for ${\tilde{V}}_{0}$ allows
to clearly stress the random
nature of the configuration patterns even for very small
changes of the lattice potential strength, whereas 
degeneracy and disorder can set in already at lower 
values of the lattice depth, depending on the tuning
of the harmonic oscillator and scattering lengths (see
below).
The degenerate states are separated by energy barriers.
The system is non--ergodic, and hysteresis should be observed:
what particular state is chosen, depends on the
exact mechanism of preparation of the system and of loading
of the mixture into the optical lattice.

The criterion for the onset of degeneracy and nonperiodic 
ground states in the bulk region around the center of the
lattice and of the trapping potential is easily identified,
by looking at the relative importance of the trapping on
site energy with respect to the on site Boson or Fermion 
interaction energy. For instance, to 
allow for the Fermionic behavior of the Boson
on site occupation numbers (either $0$ or $1$) one must
require that the energy is lower having one Boson at the
edge of the bulk central region rather than having it 
sitting on top of another Boson at the center of the 
lattice: 
\begin{equation}
{\tilde{U}}_{BB} > {\tilde{V}}_{B}
\left( i = (N_{B} + N_{F})/2 \right) \; .
\label{condition1}
\end{equation}
The analogous condition for the Bose--Fermi on site 
interaction is:
\begin{equation}
{\tilde{U}}_{BF} > {\tilde{V}}_{B/F}
\left( i = (N_{B} + N_{F})/2 \right) \; .
\label{condition2}
\end{equation}

For smaller values of $\tilde V_{0}$, the Boson hopping contribution 
will become more and more important. A representative situation of 
this intermediate regime is 
depicted in FIGs.\ \ref{Bdensityrep} and \ref{Fdensityrep}: here, the 
repulsion between Bosons and Fermions is strong enough to allow for
phase separation, while the non--negligible hopping terms still
favor configurations where Bosons have Bosons as nearest neighbors.
The transition to degeneracy and disorder, 
exact in the limit of infinite lattice depth, is a novel and
peculiar feature of Bose--Fermi mixtures and it should hold
in general for any 
multicomponent Bose and/or Fermi dilute atomic system loaded  
in a deep optical lattice at zero temperature, provided that 
intercomponent interactions are repulsive and the on site
confining potentials coincide for the different components. 
It clearly cannot take place in a single--component 
system, say a pure single--component Bose gas, where only
a SF--MI transition occurs \cite{JacBru98}.
The rather complex and rich interplay between ordered and 
disordered configurations of Bose--Fermi mixtures in very deep 
optical lattices will be considered in more detail elsewhere. 

\section{Summary and Outlook}
\label{Outlook}
In conclusion, we have studied the zero--temperature
properties of a mixture of weakly interacting gases
of neutral Bosonic and Fermionic atoms loaded in 
one--dimensional optical lattices and confined by
harmonic trapping potentials.
We have derived a single--band Bose--Fermi
Hubbard Hamiltonian, and performed some mean
field studies of the zero--temperature phase
diagram. We have considered the case of a quasi--free
Fermion sea acting on the Bosons, which have been
treated in their full dynamical range.
We have always worked in the approximation
of $s$--wave Boson--Boson and Boson--Fermion
contact interactions. According to the possible
different combinations of intraspecies 
and interspecies attractive and repulsive interactions,
the system displays a rich phase structure,
including the onset of a SF--MI transition in the
Boson sector, and a simultaneous transition to
demixing in the Boson--Fermion sector.
The optical lattice potential plays a crucial
role, allowing to tune the system into regimes
of strong Boson--Boson and Boson--Fermion couplings
as the lattice depth is increased.
For very deep lattices the system displays a 
remarkable transition to a multiply degenerate phase in
which all possible permutations of configurations
with one Bosonic or Fermionic atom per site are
legitimate ground states. The transition is
related with breaking of the lattice mirror 
symmetry for very large values of the lattice
depth. This peculiar disordered pattern of
degenerate ground--state configurations 
separated by very large barriers is somehow
reminiscent of the behavior of classical 
disordered systems like glasses and spin
glasses, but it takes place in a quantum
system at zero temperature.

The setting that has been investigated in detail 
in the present paper
can be extended in various ways. 
Certainly a larger number of Bosons and Fermions have
to be considered in order to obtain a more realistic
description of the system. While our previous analytical
findings are applicable to any numbers of atoms,
in order to extend the numerical calculations to larger numbers 
more powerful numerical methods have to be introduced. So far,
Monte Carlo simulations with a fairly large number of particles 
have been carried out only for an inhomogeneous Bose--Hubbard model
\cite{Batrouni}.
The authors have also speculated that the qualitative phase
diagram does not depend on the dimensionality of the system. 

In order to extend the work presented in this paper
to the case of interacting Fermions, one may either
allow for different Fermionic species in magnetic traps 
or for spin--unpolarized identical Fermions in optical
traps. Exchange--correlation effects, that are already
included in the Gutzwiller Ansatz for the Bosons,
will then become important for the Fermions as well \cite{Lieb}.
In this way the Bose--Fermi interaction has to be consistently
incorporated beyond the mean--field level. In the present
paper the mean--field treatment of the Bose--Fermi interaction 
is consistent, since we only included the Fermions
in a mean--field and LDA prescription. 

At the opposite extreme, one can consider the Fermions 
to be static impurities for the Bosons, as
very recently discussed by Vignolo and collaborators
\cite{Vignolo}. Finally,
after the first version of the present paper
had been submitted, a preprint appeared on
the induced Boson--Boson
interaction due to the Fermions for mixtures
loaded in a $2$--$D$ lattice \cite{Blatter}.
   
Besides these fundamental theoretical aspects
related to the theory of quantum phase transitions
and the statistical mechanics of complex systems, 
ultracold Bose--Fermi mixtures in 
an optical lattice qualify for potential
applications in the physics of quantum information.
As with systems involving either Bosons or
Fermions that have been studied so far 
\cite{JakBriCirGarZol98,Deutsch,PacKni03,GarCir03,DuaDemLuk03},
mixtures could be used for the preparation of 
multi-particle entangled states
\cite{DuaDemLuk03} such as cluster states or certain instances
of graph states \cite{Briegel}, as well as
for the implementation of quantum gates.
With Bosons and Fermions serving two different purposes,
Bose-Fermi mixtures could in fact allow for new 
possibilities of quantum information processing
in optical lattices. The Fermions would be suitable for
storage of quantum information due 
to their non--interacting behavior, whereas the Bosons 
could be used to let the systems interact and perform
operations. 

In turn, the study of such coupled quantum systems exhibiting
collective phenomena with the methods of the
theory of multi-particle entanglement
is an attractive investigation in its own right.
The aim here is not to use the coupled system to
prepare strongly entangled systems that form the 
starting point for applications in quantum 
information processing. Instead, the motivation of such 
investigations is to go beyond conventional 
methods to characterize the natural 
correlations present in the distributed system 
at zero temperature \cite{Osborne,Vidal,Chain,Zanardi}. 

The quantitative theory of 
entanglement that abstracts from the actual
physical realization of a Bose--Fermi Hubbard model
could provide the tools to understand how global 
properties emerge here from quantum correlations 
between the elementary constituents.

\section{Acknowledgments}

We thank Kai Bongs and Giovanni Modugno for
very useful discussions.
A. A. and J. E. thank the DFG and the ESF
for financial support.
F. I. thanks the INFM for financial support
as well as COSLAB and BEC2000+ ESF programs.


\begin{thebibliography}{99}

\bibitem{Greiner1} 
M. Greiner, O. Mandel, T. Esslinger, T. W. H{\"a}nsch,
and I. Bloch, Nature {\bf 415}, 39 (2002).

\bibitem{Greiner2} 
M. Greiner, I. Bloch, O. Mandel, T. W. H{\"a}nsch, and T. Esslinger,
Phys. Rev. Lett. {\bf 87}, 160405 (2001).

\bibitem{OrzTuc01}
C. Orzel, A. K. Tuchman, M. L. Fenselau, M. Yasuda, 
and M. A. Kasevich, 
Science {\bf 291}, 2386 (2001). 
							 
\bibitem{Anglin} J. R.  Anglin, and W. Ketterle,
Nature {\bf 416}, 211 (2002).

\bibitem{Jessen1} P. S. Jessen, and I. H. Deutsch,
Adv. At. Mol. Opt. Phys. {\bf 37}, 95 (1996).

\bibitem{Sachdev} S. Sachdev, {\it Quantum Phase
Transitions} (Cambridge University Press, Cambridge, 
1999).

\bibitem{JacBru98}
D. Jaksch, C. Bruder, J. I. Cirac, C. W. Gardiner, and P. Zoller,   
Phys. Rev. Lett. {\bf 81}, 3108 (1998).

\bibitem{OosStr02}
D. van Oosten, P. van der Straten, and H. T. C. Stoof,
Phys. Rev. A {\bf 63}, 053601 (2001).

\bibitem{RuoDun02}
J. Ruostekoski, G. V. Dunne, and J. Javanainen,
Phys. Rev. Lett. {\bf 88}, 180401 (2002).

\bibitem{Hofstetter02} 
W. Hofstetter, J. I. Cirac, P. Zoller, E. Demler, and M. D. Lukin,
Phys. Rev. Lett. {\bf 89}, 220407 (2002).

\bibitem{Pared02} 
B. Paredes and J. I. Cirac, 
Phys. Rev. Lett. {\bf 90}, 150402 (2003).

\bibitem{Recati03} 
A. Recati, P. O. Fedichev, W. Zwerger, and P. Zoller,
Phys. Rev. Lett. {\bf 90}, 020401 (2003).

\bibitem{Zwerger03} 
H. P. B\"{u}chler, G. Blatter, and W. Zwerger,
Phys. Rev. Lett. {\bf 90}, 130401 (2003).

\bibitem{Kerman} 
A. J. Kerman, V. Vuletic, C. Chin, and S. Chu,
Phys. Rev. Lett. {\bf 84}, 439 (2000).

\bibitem{Jessen2} P. S. Jessen, D. L. Haycock, G. Klose,
G. A. Smith, I. H. Deutsch, and G. K. Brennen,
Quant. Inf. Comp. {\bf 1}, 20 (2001).

\bibitem{DuaDemLuk03}
L.--M. Duan, E. Demler, and M. D. Lukin,
LANL Preprint cond--mat/0210564 (2002).

\bibitem{Deutsch} I. H. Deutsch, G. K. Brennen, and
P. S. Jessen, Fortsc. der Physik {\bf 48} 925 (2000). 

\bibitem{JakBriCirGarZol98}
D. Jaksch, H.--J. Briegel, J. I. Cirac, C. W. Gardiner, 
and P. Zoller, Phys. Rev. Lett. {\bf 82}, 1975 (1999).
	
\bibitem{GarCir03}	
J. J. Garcia--Ripoll and J. I. Cirac,
Phys. Rev. Lett. {\bf 90}, 127902 (2003).

\bibitem{DorFed02}
U. Dorner, P. Fedichev, D. Jaksch, M. Lewenstein, and P. Zoller,
LANL Preprint
quant-ph/0212039 (2002).

\bibitem{PacKni03}	
J. Pachos and P. L. Knight, LANL
Preprint quant--ph/0301084 (2003).

\bibitem{Fisher} M. P. A. Fisher, P. B. Weichman, G. Grinstein,
and D. S. Fisher, Phys. Rev. B {\bf 40}, 546 (1989).

\bibitem{Inpreparation} For a detailed study of the effects
of inhomogeneities in Boson--Fermion lattice models
cfr: A. Albus, M. Cramer, J. Eisert, and F. Illuminati, 
in preparation. 

\bibitem{GiovanniModugno} G. Modugno, F. Ferlaino, R. Heidemann, 
G. Roati, and M. Inguscio, LANL Preprint cond--mat/0304242 (2003).

\bibitem{Uns} A. P. Albus, S. A. Gardiner, F. Illuminati,
and M. Wilkens, Phys. Rev. A {\bf 65}, 053607 (2002).

\bibitem{Giorgini} L. Viverit and S. Giorgini, Phys. Rev.
A {\bf 66}, 063604 (2002). 

\bibitem{CheWu03}
G.-H. Chen and Y. S. Wu,
Phys. Rev. A {\bf 67}, 013606 (2003).

\bibitem{Viverit}
L. Viverit, C. J. Pethick, and H. Smith,
Phys. Rev. A {\bf 61}, 053605 (2000).

\bibitem{Freericks} J. K. Freericks and H. Monien, Europhys.
Lett. {\bf 26}, 545 (1994).

\bibitem{SheKri93}
K. Sheshadri, H. R. Krishnamurthy, R. Pandit, and T. V. Ramarishnan,
Europhys. Lett. {\bf 22}, 257 (1993).

\bibitem{KrauCaf91}
W. Krauth, M. Caffarel, and J.-P. Bouchaud,
Phys. Rev. B {\bf 45}, 3137 (1991).

\bibitem{KirGelVec83}
S. Kirkpatrick, C. D. Gelatt, and M. P. Vecchi,
Science {\bf 220}, 671 (1983).
		  	  		  
\bibitem{WahWan99}
B. W. Wah and T. Wang,
{\it Simulated Annealing with Asymptotic Convergence 
for Nonlinear Constrained Global Optimization}, 
Proc. Principles and Practice of Constraint 
Programming (Springer, Heidelberg, 1999).
		  
\bibitem{NelMea65}
J. A. Nelder and R. Mead,
Comp. Journal {\bf 7}, 308 (1965).

\bibitem{Roth}
R. Roth, Phys. Rev. A {\bf 66}, 013614 (2002).

\bibitem{Cazalilla} M. A. Cazalilla and A. F. Ho,
LANL Preprint cond--mat/0303550 (2003).

\bibitem{Uns2} A. P. Albus, F. Illuminati, and
M. Wilkens, LANL Preprint cond--mat/0211060, and 
Phys. Rev. A {\bf 67} (2003), in press.		

\bibitem{Batrouni} 
G. G. Batrouni, V. Rousseau, R. T. Scalettar, 
M. Rigol, A. Muramatsu, P. J. H. Denteneer,
and  M. Troyer, Phys. Rev. Lett. {\bf 89}, 117203 (2002). 

\bibitem{Lieb} E. H. Lieb and F. Y. Wu,
Phys. Rev. Lett. {\bf 20}, 1445 (1968);
{\it ibidem} {\bf 21}, 192 (1968). 

\bibitem{Vignolo} P. Vignolo, Z. Akdeniz, and M. P. Tosi,
LANL Preprint cond--mat/0304104 (2003).

\bibitem{Blatter} H. P. B\"{u}chler and G. Blatter,
LANL Preprint cond--mat/0304534 (2003).
 
\bibitem{Briegel}
W. D\"{u}r and H.-J. Briegel,
Phys. Rev. Lett. {\bf 90}, 067901 (2003).

\bibitem{Osborne}
T. J. Osborne and M. A. Nielsen, 
Phys. Rev. A {\bf 66}, 032110 (2002).

\bibitem{Vidal}
J. I. Latorre, E. Rico, and G. Vidal,
LANL Preprint quant--ph/0304098 (2003).

\bibitem{Chain}
K. Audenaert, J. Eisert, M. B. Plenio, and R. F. Werner,
Phys. Rev. A {\bf 66}, 042327 (2002).

\bibitem{Zanardi}
P. Giorda and P. Zanardi,  
LANL Preprint quant--ph/0304151 (2003).

\end{thebibliography}
\end{document}